# Neutron Reflectometer with Polarization Option at the Budapest Neutron Centre

**L. Bottyán[1], D.G. Merkel[1], B. Nagy[1], J. Major[1,2]**

[1]*KFKI Research Institute for Particle and Nuclear Physics, H-1525 Budapest, P.O.B. 49 Hungary*

[2]*Max-Planck-Institut für Intelligente Systeme (formerly Max-Planck-Institut für Metallforschung), Heisenbergstr. 3, D-70569 Stuttgart, Germany*

## INTRODUCTION

The ever increasing need for product advancement and miniaturization keeps thin film assemblies, membranes, magnetic and non-magnetic multilayer and patterned heterostructures in the limelight of material science and technological development. A number of thin film and surface characterization methods have emerged recently to meet the new challenges. The increased interest in magnetic thin film analytical instruments – mainly triggered by the discovery of the giant magnetoresistance and related phenomena[1] – resulted in a boom of PNR studies as well as of the construction of a number of new neutron reflectometers with polarization option at neutron sources all over the world. Here we report on the design, construction and operation parameters and first example uses of the "Grazing Incidence Neutron Apparatus" (GINA) a recently installed neutron reflectometer at the Budapest Neutron Centre (BNC), Hungary.

# GENERAL OVERVIEW OF THE INSTRUMENT

The GINA reflectometer is a constant-energy angle-dispersive, vertical-sample instrument[2]. The setup is displayed in Figure 1.

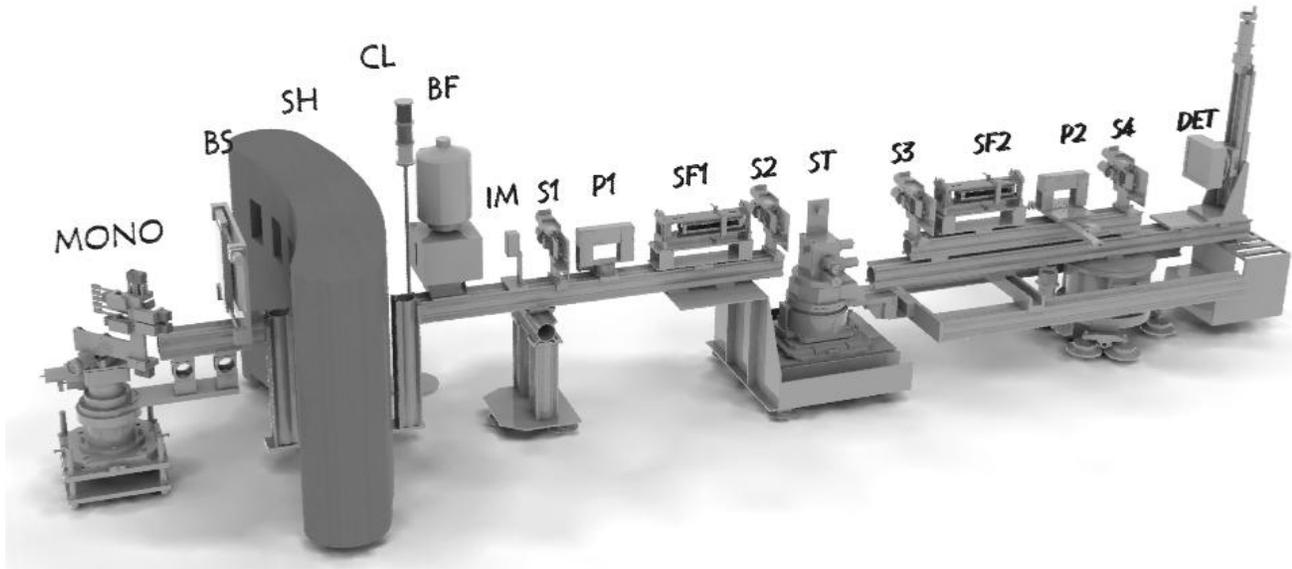

Figure 1 The layout of the GINA neutron reflectometer. The MONO assembly is mounted behind the concrete shielding (SH) on a turntable connected to the optical bench (B1) supporting the beam shutter (BS), (latter monitored by semaphore control light, CL), the beryllium filter (BF), the intensity monitor detector (IM), slit S1, polarizer P1, adiabatic RF spin flipper (SF1) and slit S2. Bench B1 is connected to the sample tower (ST) The optical bench B2 is connected to the turn-table underneath ST1 and supports the slit (S3), the second spin flipper (SF2), the spin analyzer P2 and optional slit S4 in front of the detector (DET).

The monochromator MONO provides neutrons with wavelengths within the range of 3.2Å – 5.7Å and $\Delta\lambda/\lambda \approx 1\%$. The polarized neutron beam is produced by using a magnetized supermirror (P1) and an adiabatic radio-frequency (RF) spin flipper[3] (SF1). The beam scattered on the sample may undergo spin analysis by an identical setup of a spin flipper and a spin analyzer (P2), and finally it is detected by a two-dimensional position sensitive neutron detector (DET). The incident intensity is monitored by a low efficiency (~0.1% at $\lambda = 4.6$Å) beam intensity monitor (IM). The components of the reflectometer are mounted on two heavy-load optical benches B1 and B2. B1 supports the beam shutter (BS), the IM, the beryllium filter (BF), the slit (S1) and the SM polarizer (P1), the adiabatic RF spin flipper (SF1) and the slit (S2). The downstream end of the bench B1 is fixed to the central sample tower ST and supports the various sample environment components (electromagnet, cryostat, etc.). The incident angle on the sample surface is set by the major (θ-) goniometer of ST. The bench B2 – the 2θ-arm of the reflectometer – supports the slit S3, the spin flipper SF2, the spin analyzer P2, and the 2D detector along with its electronics and dedicated control PC mounted underneath. The slit S4 in front of the detector is optionally used when data collection is restricted to specularly scattered neutrons. The 2θ-motion is driven by a wheel running on the marble surface while the corresponding air pads are activated. The wavelength may be changed by manually rotating the entire GINA setup around the turntable while air pads are activated and both arms float over the marble floor. The available wavelengths are restricted at present to 3.2, 3.9, 4.6, 5.2 and 5.7 Å by the respective channels through the cylindrical concrete shielding (SH) around the monochromator assembly.

The monochromator is located in a gap of the curved Ni(Mo)/Ti SM guide 19 meters downstream the cold source, and comprises five highly oriented pyrolytic graphite crystals on small motorized 2-circle cradles for horizontal alignment and vertical focusing. Vertical focusing of the beam to the sample position doubled the intensity reflected by a 20×20 mm$^2$ sample at grazing incidence as compared to the non-focused case of parallel graphite crystals. Higher harmonics intensity is efficiently filtered by a Be block. The transmission of the filter is 41% and 87% for λ=4.6Å, without and with liquid nitrogen cooling, respectively.

Polarized neutrons are produced by an Fe-Co/Si magnetic SM placed in an in-plane vertical magnetic field of 30 mT in transmission geometry (P1 in Figure 1). Spin analysis of the specularly reflected beam is performed by a single magnetic SM analyzer (P2) of identical construction with P1. In order to decrease the scattering of neutrons by the beam-line components, adiabatic RF spin flippers[3] are installed. The flipper coil is placed in a longitudinal gradient field of 20 – 40 mT/m, with a center field of 5.6 mT produced by two iron plates energized by permanent magnets upstream and shunted downstream. The flipper coil is part of a resonant circuit, with typical values of effective RF current and bandwidth of 4 A and 4.5 kHz at the resonance frequency of 166 kHz.

Fine definition of the beam is maintained by the four slits with cadmium blades. The blades are operated with a precision of 0.05 mm and are covered with B$_4$C-containing material. Slit S1 defines the beam on the polarizer P1 to decrease the divergence thus increasing the polarization ratio. Slit S2 decreases the beam divergence on the sample and absorbs the neutrons scattered by the polarizer. With these optical elements the setup exhibits a relative Q-resolution of 10 to 2% for the available Q-range of 0.005 to ~0.25 Å$^{-1}$.

TABLE I. Operation parameters of GINA

| Parameter | Range |
|---|---|
| Wavelength | 3.9÷5.1 Å in five steps |
| Present wavelength | 4.6 Å |
| Max. scattering angle | ≥θ=35° |
| Angular resolution (Δθ) | 0.003° |
| Δλ/λ | ~1% |
| Background level | 0.01 cps cm$^{-2}$ |
| Detector | 2D PSD, 200×200 mm$^2$ |
| Detector spatial resolution | 1.6×1.6 mm$^2$ |
| Neutron flux at the monochromator position | 4×10$^5$ n×cm$^{-2}$×s$^{-1}$ |
| Background reflectivity | < 7×10$^{-5}$ |
| Aggregate polarization efficiency | 0.895 |

Scattered neutrons are registered by a delay line type multi-wire proportional chamber of 200×200 mm$^2$ active area and spatial resolution of 1.6 mm (FWHM) filled with $^3$He / $CF_4$ gas mixture of 2.5/3 bar partial pressures. The detector is encased in a boron-containing shielding for background suppression. A DASY TDC module (produced by ESRF, Grenoble) is installed in a slot of the PC dedicated exclusively to the detector data acquisition and mounted on the $2\theta$-arm of the reflectometer. If no spin analysis is required, for further background suppression, an evacuated flight tube is mounted along the entire length of the $2\theta$ arm. Encasing the analyzer P2 and flipper SF2 in a vacuum vessel is a plan for the near future.

**MEASUREMENT CONTROL**

The GINA hardware and its control software are designed for maximum flexiblility and remote controllability. In its full configuration, GINA comprises more than 30 stepping motors. All motions are remotely controllable. Certain critical motions, such as $\theta$- and $2\theta$-angles and precision slit positions are absolute or relative encoder-controlled. Hardware control is maintained via a custom made unit built around a USB multi-function data acquisition module to control the air compressor, the air pads, the temperature of the Be-filter, the beam shutter and control lights, the beam intensity monitor and various modular DC power supplies. The high voltage power supplies, the linear amplifiers, the discriminators and the ratemeters are of NIM standard. The control PC directly communicates with the detector PC via ethernet and with the indexer modules of the motion control units as well as with the temperature controller via RS232 under the supervision of GINASoft written in LabView 2009 for MS Windows. The program user interface supports various component alignments and scan modes via macros as well as polarization and the sample environment control (flipper current and frequency, sample temperature, magnet current, etc.) to perform remotely. The user interface is highly configurable. A screenshot of two windows of the GINASoft is displayed in Figure 2.

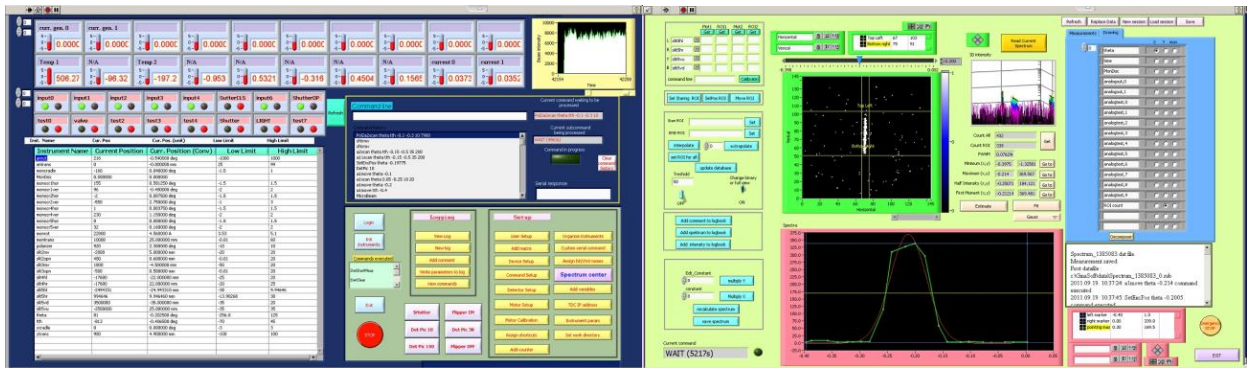

Figure 2 Two screenshots of the GINASoft control program. For further details see: www.bnc.hu/.

2D detector pictures and reduced reflectivity data are efficiently viewed and manipulated during the data acquisition. Collected and manipulated data as well as extended log information are saved in a clearly structured database format. Human control is facilitated by a web camera. Using remote desktop option, most operations can be performed remotely via internet from outside the experimental hall or even from a distant continent.

**SAMPLE ENVIRONMENT**

GINA is dedicated to reflectometry of magnetic heterostructures. For studies of magnetism, vital environmental parameters are (low) temperature and (occasionally high) external magnetic fields. The sample mounting depends on the sample environment. For room temperature reflectivity measurements the sample is held in position by vacuum. Two cradles and two perpendicular translators position the sample in the vertical plane and set the sample surface orientation. A closed cycle cryostat can be mounted on the sample tower ST with or without the electromagnet. The sample temperature can be varied in the 9 to 300 K range. At GINA an air-cooled electromagnet is available, which generates magnetic fields up to 0.55 T for the pole distance of 40 mm that accommodates the 1.5" diameter cryostat housing. The optional water-cooled air core coil pair provides fields up to approx. 35 mT.

**EXAMPLE REFLECTOGRAM**

An example reflectogram is chosen to highlight the polarized specular reflectivity performance of GINA. The sample was a 20×20 mm$^2$ isotope-periodic multilayer prepared by molecular beam epitaxy in our group[1] with a MgO(001)/[Ni(15nm)/$^{62}$Ni(5nm)]$_5$ nominal layer structure. Measured $R^+$, $R^-$ reflectivities and the derived quantity, $(R^+ - R^-)/(R^+ + R^-)$ are displayed in Figure 3 in panels a), b) and c), respectively.

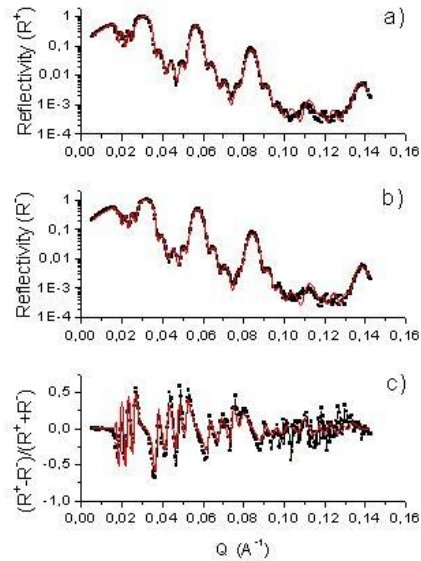

Figure 3 Measured $R^+$, $R^-$ and derived $(R^+-R^-)/(R^++R^-)$ specular reflectivities of MgO(001)/[Ni(15nm)/$^{62}$Ni(5nm)]$_5$ measured at GINA in polarized mode.

Data collection time was 56 hours. The *simultaneous* fit to $R^+$ and $R^-$ by FitSuite[4] using nominal SLDs and collinear magnetization of 3 kG is displayed in full lines.

**SUMMARY**

We have shown that GINA, a dance-floor-type, vertical sample, constant energy angle-dispersive neutron reflectometer, recently installed at the Budapest Neutron Centre is a versatile instrument. Sample environments comprise a closed cycle cryostat from 9 K to 300 K, an electromagnet up to 0.6 T or air core coils of 35 mT, spin polarization and polarization analysis by single supermirrors. Detection of diffuse scattering is facilitated by a two-dimensional position-sensitive neutron detector. Reflectivity ranges above four orders of magnitude have been measured with improvements planned. Further developments including an environmental cell for membrane studies, a supermirror fan analyzer and further background suppression elements are underway. The GINA reflectometer is open for Hungarian and international users throughout the year. Information concerning proposal submissions can be found at

---

[1] MBE preparation and characterization of metallic multilayers for external users – including deposition of various stable isotope layers ($^{57}$Fe, $^{62}$Ni, etc.) – are offered as a service.

www.bnc.hu. Since the instrument is operated by a metallic thin film group, MBE preparation and characterization of metallic multilayers including deposition of various stable isotope layers ($^{57}$Fe, $^{62}$Ni, etc.) can also be arranged.

**ACKNOWLEDGEMENT**

The GINA team is grateful to Prof. H. Dosch, former director of Max-Planck-Institut für Metallforschung for his continued interest in the GINA project and for the transfer of a number of components of EVA, a former neutron reflectometer operated by the Max-Planck-Institut für Metallforschung at the Institut Laue-Langevin, Grenoble, France. This work was supported by the National Office for Research and Technology of Hungary and the Hungarian National Science Fund (OTKA) under contracts NAP-VENEUS and K 62272, respectively.

---

[1] Review articles in "*Nanomagnetism and Spintronics*" Ed. T. Shinjo, ISBN: 978-0-444-53114-8, Elsevier (2009)

[2] L. Bottyán, D.G. Merkel, B. Nagy, Sz. Sajti, L. Deák, G. Endrőczi, J. Füzi, A. V. Petrenko, J. Major, submitted to Rev. Sci. Instr. (2011)

[3] S.V. Grigoriev, A.I. Okorokov and V.V. Runov, *Nucl. Instrum. and Meth.* A 384, 451 (1997)

[4] www.fs.kfki.hu